\documentclass[a4paper]{jpconf}
\usepackage{graphicx}
\usepackage{iopams}

\begin{document}
\title{Incoherent bremsstrahlung in flat and bent crystals}

\author{N F Shul'ga$^1$, V V Syshchenko$^2$ and A I Tarnovsky$^2$}

\address{$^1$ Akhiezer Institute for Theoretical Physics of the NSC
``KIPT'', Akademicheskaya Street, 1, Kharkov 61108, Ukraine}

\address{$^2$ Belgorod State University, Pobedy Street, 85, Belgorod 308015, Russian Federation}

\ead{shulga@kipt.kharkov.ua, syshch@bsu.edu.ru, syshch@yandex.ru}

\begin{abstract}
Incoherent bremsstrahlung by high-energy particles in crystal is
due to the thermal spread of atoms in relation to their
equilibrium positions in the lattice. The simulation procedure
developed earlier for the incoherent radiation is applied to the
case of the electrons and positrons motion in the sinusoidally
bent crystal. The results of simulation are in agreement with the
data of recent experiments carried out at the Mainz Microtron
MAMI. The possibility of use of the sinusoidally bent crystals as
undulators is discussed.
\end{abstract}

\section{Introduction}
The bremsstrahlung cross section for relativistic electrons in a
crystal is split into the sum of the coherent part (due to the
spatial periodicity of the atoms' arrangement in the crystal) and
the incoherent one (due to the thermal motion of atoms in the
crystal) \cite{TM, AhSh, Ugg}. Although the spectrum of incoherent
radiation in crystal is similar to one in amorphous medium, the
incoherent radiation intensity can demonstrate substantial
dependence on the crystal orientation due to the electrons' flux
redistribution in the crystal (channeling etc.). The simulation
based on the semiclassical description of the radiation process
\cite{Sh2, Sh3, Sh4} confirms that viewpoint. The results of
simulation are in a good agreement with the corresponding early
\cite{Sanin} and recent \cite{Backe1} experimental data.

Here we present the results of simulation using the improved
procedure taking into account the crystal deformations. The
simulation was carried out under the conditions of the recent
experiment performed at the Mainz Microtron MAMI \cite{Backe2} to
explore the radiation emission from periodically bent crystal. The
possibility of application of such crystals as undulators is
discussed during last years \cite{Backe2, Tabrizi, Backe3}.

For the reader convenience, in the next section we outline some
theoretical ideas of our approach.

\section{Computation method}

Let us consider the high energy electron incidence on the atomic
string in the crystal. The two-dimensional multiple scattering
angle $\boldsymbol\vartheta$ is equal to the sum of individual
scattering angles on the atoms:
\begin{equation}
\boldsymbol\vartheta = \sum_n \boldsymbol\vartheta
(\boldsymbol\rho_n),  \label{sst:eq1}
\end{equation}
where $\boldsymbol\rho_n$ is the impact parameter of the collision
with the \emph{n}-th atom of the crystal. The mean square of the
multiple scattering angle (averaged over the thermal vibrations of
atoms) can be expressed as the sum of two blocks of terms:
\begin{equation}
\left<\left| \sum_n \boldsymbol\vartheta (\boldsymbol\rho_n)
\right|^2\right> = \sum_{n,m} \left< \boldsymbol\vartheta
(\boldsymbol\rho_n) \right>\left< \boldsymbol\vartheta
(\boldsymbol\rho_m)\right> + \sum_n \left\{\left<
\boldsymbol\vartheta (\boldsymbol\rho_n)^2\right> - \left<
\boldsymbol\vartheta (\boldsymbol\rho_n)\right>^2 \right\}.
\label{sst:eq2}
\end{equation}
The first block describes the coherent scattering which can be
interpreted as a motion in the uniform string potential
\cite{AhSh, Ugg, AhSh2}. The second one describes the incoherent
scattering of the electron on the thermal vibrations of the
lattice atoms.

The bremsstrahlung spectrum of the electrons passing through the
crystal also can be expressed as the sum of the coherent and
incoherent contributions \cite{TM, AhSh, Ugg}. For the electrons
of the energy $\varepsilon\sim 1$ GeV the main contribution of the
coherent effect is made to the soft range of the spectrum (the
photon energy $\hbar\omega$ is less or of the order of dozens of
MeVs). In the medium and hard ranges of the spectrum the
incoherent radiation is predominant.

The spectral density of the incoherent radiation from the
individual electron moving on the given trajectory is described by
the formula \cite{Sh3, Sh4}
\begin{equation}
\left( {d\mathcal E\over d\omega} \right)_{incoh} =
{2e^2\varepsilon (\varepsilon - \hbar\omega )\over 3\pi m^2c^5}
\left\{ 1 + {3\over 4} {(\hbar\omega)^2\over
\varepsilon(\varepsilon - \hbar\omega )} \right\} \sum_n
\left\{\left< \boldsymbol\vartheta (\boldsymbol\rho_n)^2\right> -
\left< \boldsymbol\vartheta (\boldsymbol\rho_n)\right>^2 \right\}
,  \label{sst:eq3}
\end{equation}
where $m$ and $e$ are the electron's mass and charge, $c$ is the
velocity of light. It is convenient to compare the incoherent
radiation intensity of the uniform electron beam in the crystal to
the corresponding intensity in the amorphous medium (described by
Bethe-Heitler formula). The ratio of these two values is equal to
\cite{Sh3, Sh4}
\begin{equation}
N_\gamma = {1\over 2\pi N R^2 \ln (mRc/\hbar)} \int d^2 \rho_0
\sum_n \left\{\left< \boldsymbol\vartheta
(\boldsymbol\rho_n)^2\right> - \left< \boldsymbol\vartheta
(\boldsymbol\rho_n)\right>^2 \right\} ,  \label{sst:eq4}
\end{equation}
where $N$ is the total number of the electron's collisions with
atoms under its motion through the crystal, $R$ is Thomas-Fermi
radius of the atom, integration over $d^2\rho_0$ means the
integration over all possible points of incidence of the electron
on the crystal in the limits of one elementary cell.

The values of the function $F(\rho) = \left< \boldsymbol\vartheta
(\boldsymbol\rho)^2\right> - \left< \boldsymbol\vartheta
(\boldsymbol\rho)\right>^2$ are determined by linear interpolation
of the values pre-calculated on the regular grid of impact
parameters. The impact parameters $\boldsymbol\rho_n$ are found
using the electron's trajectory obtained by numerical integration
of the equation of motion in the field of the set of parallel
uniform strings. The influence of the incoherent scattering by the
thermal vibrations of atoms on the electron's trajectory can be
taken into account by adding to each component of the electron's
transverse velocity the random value with the dispersion
$c\sqrt{\left\{ \left< \boldsymbol\vartheta
(\boldsymbol\rho)^2\right> - \left< \boldsymbol\vartheta
(\boldsymbol\rho)\right>^2 \right\}/2}$ after each collision. For
the further computational details see \cite{Sh3, Sh4}.

\begin{figure}
\includegraphics[width=\textwidth]{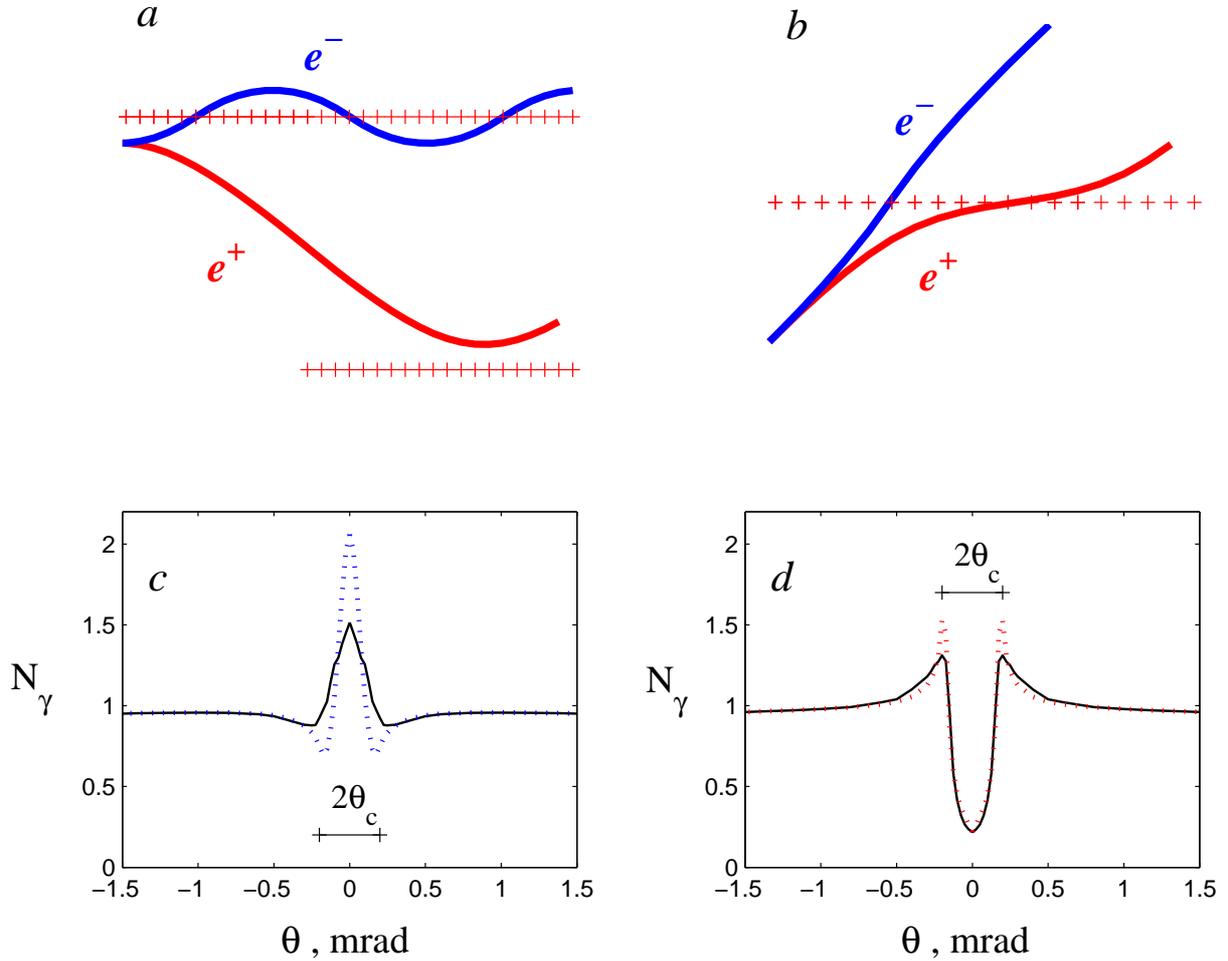}
\caption{{\bf(a)} Typical trajectories of the electrons and
positrons under planar channeling. Pluses mark the positions of
atomic strings (perpendicular to the figure plane) forming the
atomic planes of the crystal. The horizontal scale of the figure
is highly compressed. {\bf (b)} The same for above-barrier motion.
{\bf (c)} Simulated incoherent bremsstrahlung intensity (in ratio
to the Bethe-Heitler intensity in amorphous medium, see
\eref{sst:eq4}) from 1 GeV electrons vs incidence angle $\theta$
to (110) plane of 30 $\mu m$ thick Si crystal \cite{Sh2}. Dotted
line corresponds to the trajectories simulated neglecting thermal
vibrations of atoms. {\bf (d)} The same for positrons}
\end{figure}

\section{Results of simulation}
The origin of the orientation dependence of the incoherent
radiation intensity is illustrated on the figure 1. When the
electron is incident to the atomic plane of the crystal under
angle $\theta$ less than some critical angle $\theta_c$, it can be
captured by the attractive potential of the plane. The finite
motion in that potential is called as the planar channeling (see,
e.g., \cite{AhSh, Ugg, AhSh2}).

Under planar channeling the electrons collide with atoms under
small impact parameters more frequently than in amorphous medium,
that leads to the increase of the incoherent bremsstrahlung
intensity; for above-barrier motion the situation is opposite. The
account of the incoherent scattering of the particles on the
thermal vibrations of the atoms leads to dechanneling and, hence,
to the smoothing of the describer orientation dependence (compare
solid and dotted lines on the figure 1 (c) and (d)).

In the present article the simulation was carried out under the
conditions of the experiment \cite{Backe2}, where the radiation
from $\varepsilon$ = 855 MeV electrons under their incidence onto
the silicon crystal with sinusoidally bent (110) planes had been
studied. The yield of photons with the energy $\hbar\omega =
\varepsilon/2$ (for which the incoherent mechanism of
bremsstrahlung is predominant) had been registered.

\begin{figure}
\includegraphics[width=\textwidth]{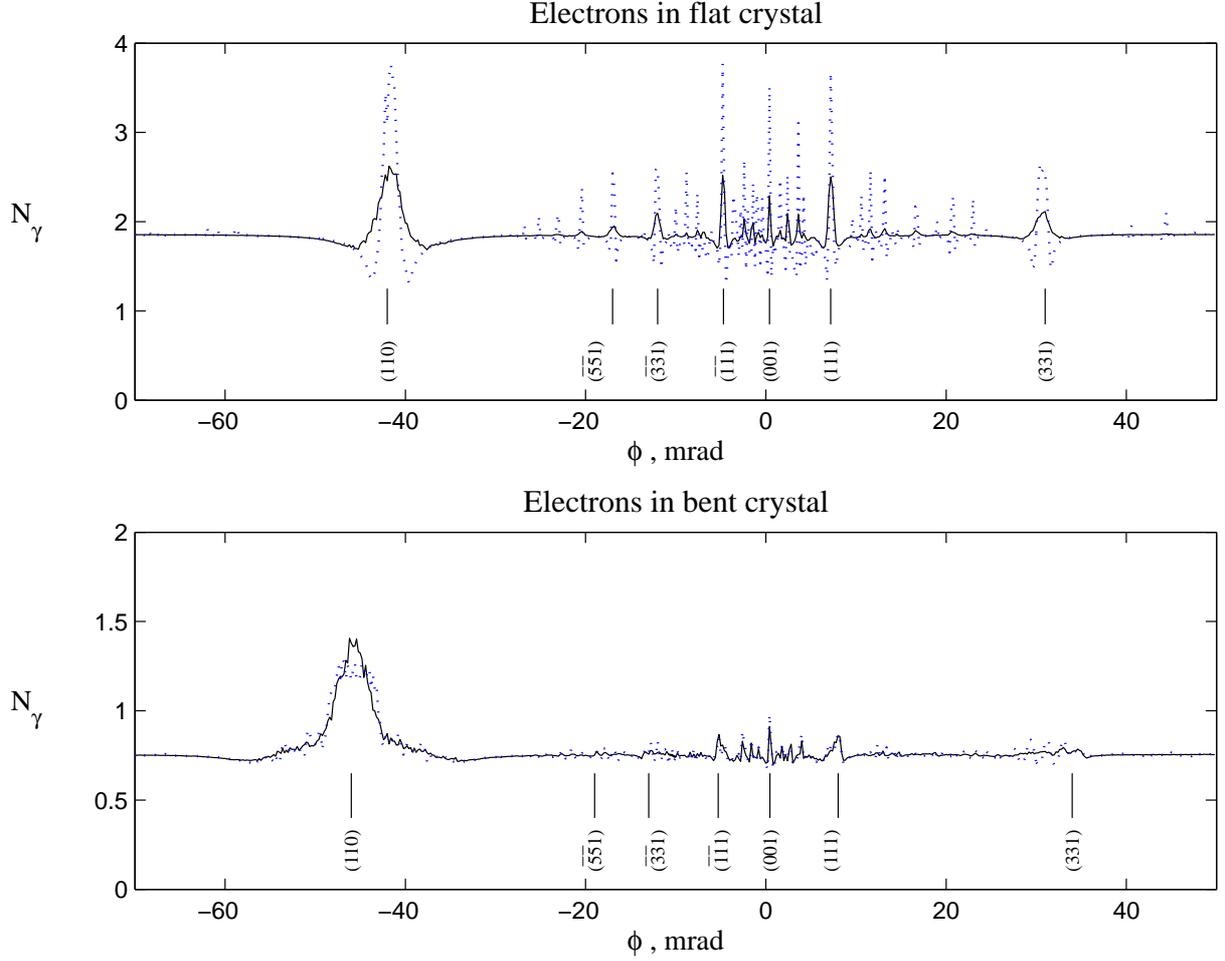}
\caption{The incoherent bremsstrahlung intensity (in ratio to
Bethe-Heitler intensity in amorphous medium) from 855 MeV
electrons in flat (upper plot) and sinusoidally bent (lower plot)
silicon crystals under scanning of the goniometric angle $\phi$
like in the experiment \cite{Backe2}}
\end{figure}

The results of simulation (figure 2) demonstrate the qualitative
agreement with the experimental data \cite{Backe2}. We can see
characteristic structures similar to one on the figure 1 (c),
generated by different crystallographic planes with the common
[1$\bar 1$0] axis.

The decrease of the radiation intensity in comparison to the
reference flat crystal permits to estimate the reduction of the
dechanneling length due to the crystal bending. We can see that
the bending of the crystallographic planes increases the
dechanneling rate so highly that the scattering on the thermal
vibrations of atoms already have no substantial influence on the
incoherent radiation intensity (compare solid and dashed curves on
the lower panel of the figure 2).

The main cause of the rapid dechanneling in the bent crystal lies
in the arising of the centrifugal addition to the planar potential
\cite{AhSh2}:
\begin{equation}
U_{eff} = U(x) - \varepsilon {x\over R_b} \ \ \ \mbox{under} \ \ \
|x|\gg R_b  \label{sst:eq5}
\end{equation}
(where $R_b$ is the bending radius) and, as a consequence, to the
reduction of the potential barriers between which the channeling
could take the place. For the crystal \cite{Backe2} the bending
radius $R_b \gtrsim 6.2\cdot 10^{-3}$ m. Assuming the planar
potential has the shape of quadratic parabola with the depth $U_0
\approx -23.5$ eV for (110) plane, the effective potential
\eref{sst:eq5} in the domain of maximal curvature of the crystal
would have the shape presented by the curve 2 on the figure 3. We
see substantial decrease of the potential wells depth comparing to
the reference flat crystal (curve 1). So, only a small part of
incident particles could be captured into the channeling regime.

\begin{figure}
\includegraphics[scale=0.54]{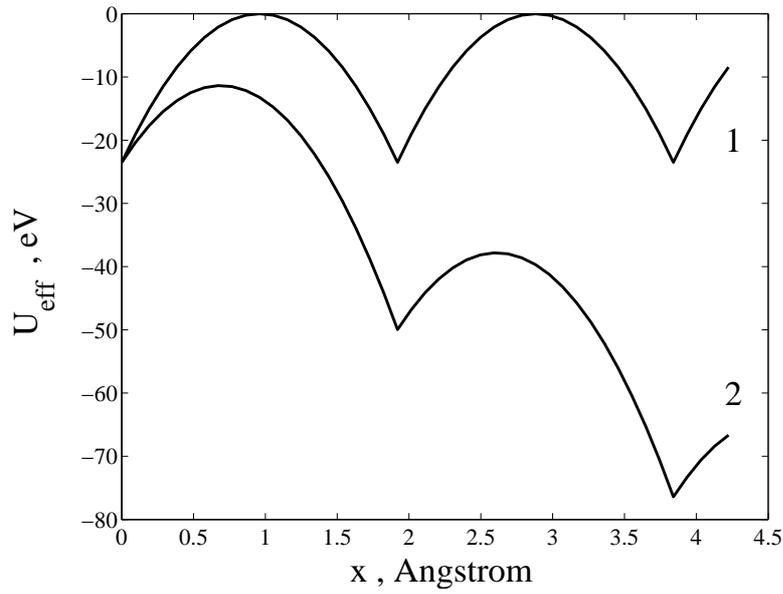}\hspace{1pc}
\begin{minipage}[b]{12.5pc}\caption{Potential energy of the electron in the planar potentials
of the flat (curve 1) and bent (curve 2) crystals.}
\end{minipage}
\end{figure}

\begin{figure}
\includegraphics[width=\textwidth]{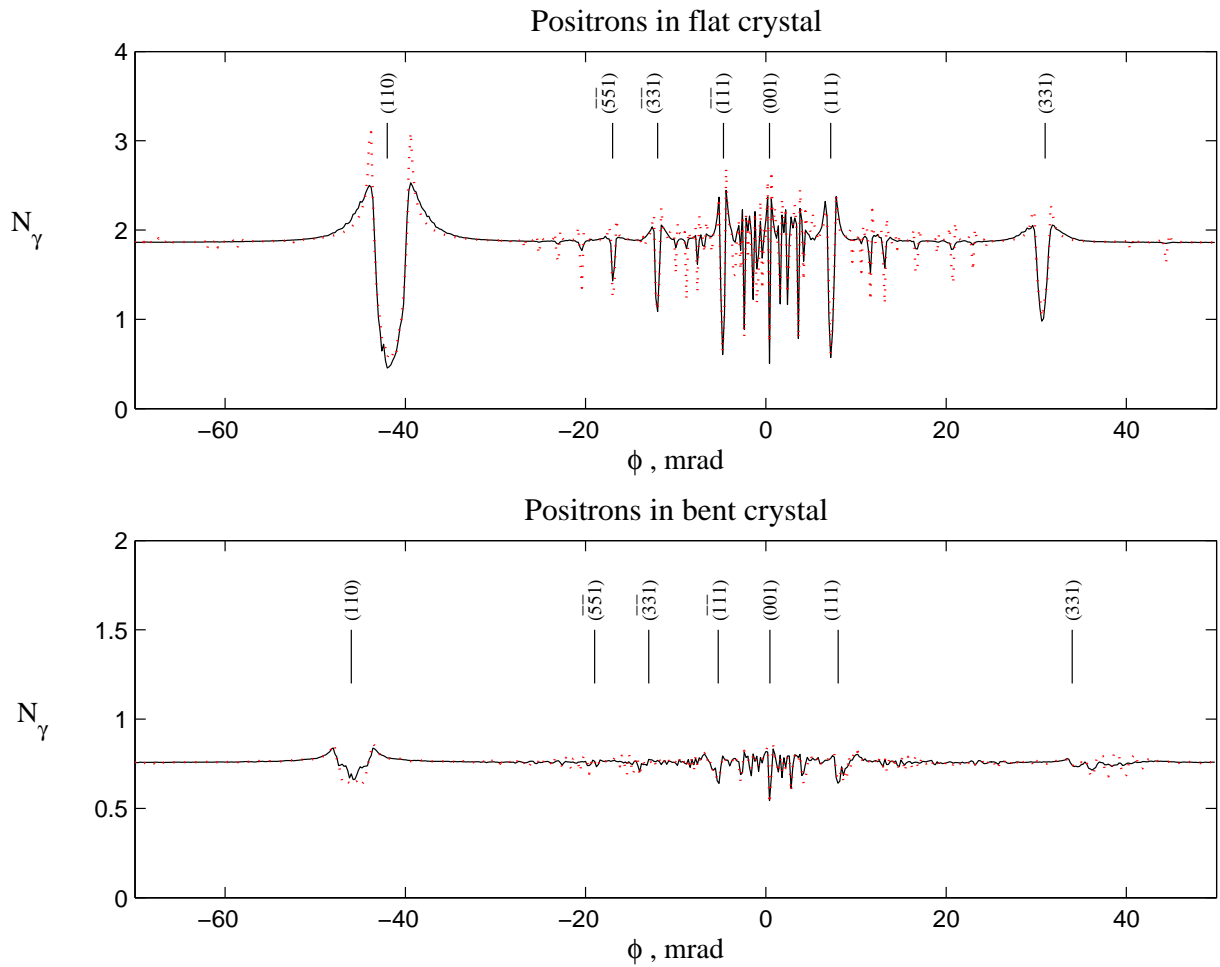}
\caption{The same as on the figure 2 for positrons.}
\end{figure}

\section{Conclusion}
The incoherent bremsstrahlung yield exceedance (deficiency) from
the electron (positron) beam in the oriented crystal in comparison
to the value predicted by Bethe-Heitler formula for the amorphous
medium can be used as an indicator of the relative amount of the
particles moving in the channeling regime.

The results of simulation demonstrate that the planar channeling
effect leads to the orientation dependence of the incoherent
bremsstrahlung intensity not only in flat, but also in bent
crystals. However, in the bent crystal the electrons rapidly leave
the planar channels. The figure 4 demonstrates that for the
positron beam instead of the electron one the dechanneling rate
would be almost the same. The effective capture of particles into
planar channels (and so the use of sinusoidally bent crystals as
undulators) is possible under changing the bending parameters to
ensure smaller curvature than in the crystal \cite{Backe2}.

\ack This work is supported in part by internal grant of Belgorod
State University and the federal program ``Academic and Teaching
Staff of innovation Russia'', the government contract
16.740.11.0147 dated 02.09.2010.


\section*{References}
\begin{thebibliography}{12}

\bibitem{TM} Ter-Mikaelyan M L 1972 {\it High-Energy Electromagnetic Processes in Condensed Media} (New
York: Wiley-Interscience)

\bibitem{AhSh} Akhiezer A I and Shul'ga N F 1996 {\it High-Energy Electrodynamics in Matter} (Amsterdam: Gordon and Breach)

\bibitem{Ugg} Uggerh{\o}j U I 2005 {\it Rev. Mod. Phys.} {\bf 77} 1131

\bibitem{Sh2} Shul'ga N F and Syshchenko V V 2005 {\it Nucl. Instr. and
Meth.} B {\bf 227} 125

\bibitem{Sh3} Shul'ga N F, Syshchenko V V and Tarnovsky A I 2008
{\it Nucl. Instr. and Meth.} B {\bf 266} 3863

\bibitem{Sh4} Shul'ga N F, Syshchenko V V and Tarnovsky A I 2010 {\it J. Phys.: Conf. Series} {\bf 236} 012027

\bibitem{Sanin} Sanin V M, Khvastunov V M, Boldyshev V F and Shul'ga N F 1992 {\it Nucl. Instr. and Meth.} B {\bf 67} 251

\bibitem{Backe1} Backe H, Kunz P, Lauth W and Rueda A 2008 {\it Nucl. Instr. and Meth.} B {\bf 266} 3835

\bibitem{Backe2} Backe H, Krambrich D, Lauth W, Lundsgaard Hansen J and Uggerh\o j U I {\it Il Nuovo Cimento (in
press)}

\bibitem{Tabrizi} Tabrizi M, Korol A V, Solov'yov A V and Greiner W 2007 {\it Phys. Rev. Lett.} {\bf 98} 164801

\bibitem{Backe3} Backe H, Lauth W, Kunz P, Rueda A, Esberg J, Kirsebom K, Lundsgaard Hansen J and Uggerh\o j U I 2010 {\it Charged and Neutral Particles
Channeling Phenomena} ed S B Dabagov and L Palumbo (Singapore:
World Scientific) p 281

\bibitem{AhSh2} Akhiezer A I, Shul'ga N F, Truten' V I, Grinenko A A and Syshchenko V V 1995 {\it Physics-Uspekhi}
{\bf 38} 1119

\end{thebibliography}
\end{document}